\documentclass[11pt]{article}
\everymath{\displaystyle}
\usepackage{amssymb}
\usepackage{amsmath}
\newtheorem{prethm}{{\bf Theorem}}

\newenvironment{thm}{\begin{prethm}{\hspace{-0.5
               em}{\bf .}}}{\end{prethm}}
\newtheorem{prelemma}{{\bf Lemma}}

\newenvironment{lemma}{\begin{prelemma}{\hspace{-0.5
               em}{\bf .}}}{\end{prelemma}}
\newtheorem{preex}{{\bf Example}}

\newtheorem{preprop}{{\bf Proposition}}

\newenvironment{prop}{\begin{preprop}{\hspace{-0.5em}{\bf .}}}{\end{preprop}}
\newtheorem{precor}{{\bf Corollary}}

\newenvironment{cor}{\begin{precor}{\hspace{-0.5
               em}{\bf .}}}{\end{precor}}
\newtheorem{preremark}{{\bf Remark}}

\newenvironment{remark}{\begin{preremark}{\hspace{-0.5
               em}{\bf.}}}{\end{preremark}}
\newtheorem{preprob}{{\bf Problem}}

\newtheorem{predefin}{{\bf Definition}}

\newtheorem{preconj}{{\bf Conjecture}}

\newtheorem{preprobb}{{\bf Problem}}

\newtheorem{prelem}{{\bf Theorem}}

\newenvironment{proof}{{\bf Proof.}\rm }{\hfill{$\Box$}}

\newtheorem{presolution}{{\bf Solution.}}

\def\newpic#1{}
\def\qed{\ifhmode\unskip\nobreak\fi\quad\ifmmode\Box\else$\Box$\fi}

\title{\vspace{0cm}\Large\bf On irreversible spread of influence in edge-weighted graphs}
\author{\large\bf Manouchehr Zaker\footnote{E-mail: mzaker@iasbs.ac.ir}
\vspace{5mm}\\
    Department of Mathematics,\\
     Institute for Advanced Studies in Basic Sciences,\\
     Zanjan 45137-66731, Iran}

    \date{}

\begin{document}
\maketitle\
\vspace*{-0.9cm}\begin{abstract}
\noindent Various kinds of spread of influence occur in real world social and virtual networks. These phenomena are formulated by activation processes and irreversible dynamic monopolies in combinatorial graphs representing the topology of the networks. In most cases the nature of influence is weighted and the spread of influence depends on the weight of edges. The ordinary formulation and results for dynamic monopolies do not work for such models. In this paper we present a graph theoretical analysis for spread of weighted influence and mention a real world example realizing the activation model with weighted influence. Then we obtain some extremal bounds and algorithmic results for activation process and dynamic monopolies in directed and undirected graphs with weighted edges.
\end{abstract}




\section{Introduction}

\noindent Various phenomena of spread of influence in social and complex networks have been the research subject of plenty of papers in recent years. Some well-known examples of these phenomena are spread of disease in populations, propagation of virus in webs of computers, adaptation of new products in populations, spread of opinions e.g.\ in elections, default contagion in banking systems, bootstrap percolation in neural networks, etc. A common formulation of such phenomena is by graph theory. The underlying networks are represented by graphs, where vertices denote the elements of the network and edges denote the links or ties between the elements. Depending on the nature of influence the graphs are directed or undirected. During the propagation of the influence, when an element of the network has taken the influence then it's called an active element. Many phenomena are progressive or irreversible in the sense that when an element becomes active in some phase of the activation process then it remains active until end of the process. In progressive phenomena it's important to explore subsets $D$ of vertices such that when $D$ is initially activated then its activation spreads to the whole network. In terms of graph theory such special subsets are called dynamic monopolies. We are ready now to introduce the formal concepts. Corresponding to any activation process there exists an assignment of thresholds $\tau: V(G)\rightarrow \Bbb{N}\cup \{0\}$ to the vertices of $G$ such that for each vertex $v$, $\tau(v)\leq deg(v)$, where $deg(v)$ is the degree of $v$ in $G$. The value $\tau(v)$ is interpreted as the level of susceptibility of the vertex $v$ in confrontation with the incoming influences in the network. In the following we first consider the models in which influence is reciprocal or bilateral. The terminology is easily generalized using directed graphs for unilateral influences. In Section 2 we introduce activation process in edge-weighted graphs.

\noindent Let $G$ be any undirected graph without loops and multiple edges. Let $\tau$ be a threshold assignment for the vertices of $G$. The discrete time dynamic process corresponding to $(G, \tau)$ is defined as follows. The process starts with an initially active subset $D$ of vertices. We denote the set of active vertices in phase (time) $i$ by $D_i$. Hence $D_0=D$. Then at any time $i+1\geq 1$, any vertex $v$ becomes active if at least $\tau(v)$ neighbors of $v$ are already active i.e. belong to $D_i$. Note that $D_0\subseteq D_1 \subseteq D_2 \subseteq \cdots$, hence the activation is irreversible. By a $\tau$-dynamic monopoly we mean any subset $D$ of the vertices of $G$ such that by starting from $D$, all the vertices of $G$ becomes active i.e. for some $i$, $V(G)=D_i$. By the size of a dynamic monopoly $D$ we mean the cardinality of $D$. If $D$ is a dynamic monopoly then there exists some $t$ and disjoint subsets $M_0, \ldots, M_t$ such that $D=M_0$ and $V(G)=M_0 \cup M_1 \cup \cdots \cup M_t$. Following \cite{Z2}, the smallest size of any dynamic monopoly of $(G,\tau)$ is denoted by $dyn_{\tau}(G)$. Dynamic monopolies in undirected graphs with various types of threshold assignments were widely studied in the literature. See e.g.\ \cite{ABW, CL3, CDPRS, DR, FKRRS, KSZ, Z, Z2}. In some applications influence in networks is unilateral or one sided. For example in the spread of opinion like voting systems, experts have significant influence on ordinary people but the converse does not hold. Another physical example is the spread of electrical excitation in neural networks. As explained in \cite{A} activation processes in neural networks are irreversible and unilateral and also depend on the threshold of nodes. Hence activations in these networks coincide with the models of this paper for directed graphs. For these models activation process and dynamic monopolies are generalized for directed graphs. In fact in directed graphs vertices are influenced by their in-neighbors. The other related concepts and quantities are defined similarly for directed graphs. Dynamic monopolies in directed graphs have been investigated in \cite{ABW, CL3, KSZ2}. Algorithmic problems concerning dynamic monopolies were also widely studied in the literature. The main decision problem is called ``Target Set Selection".

\noindent {\bf Name:} Target Set Selection (TSS)
\newline \noindent {\bf Instance:} An undirected graph $G$ with an assignment of thresholds $\tau$ and an integer $k$.
\newline \noindent {\bf Question:} Does there exist a dynamic monopoly in $G$ with at most $k$ vertices?

\noindent Chen \cite{C} has obtained interesting hardness and inapproximability results for TSS. For the other algorithmic results see \cite{BHLN, CDPRS, C, DR, NNUW}. In topics concerning graph activation processes, simple and strict majority threshold assignments are mostly studied because many spread phenomena such as spread of opinion in voting systems, adaptation of new products in populations and fault propagation in distributed computing correspond to these special types of assignments. In simple majority assignment we have $\tau(v)=deg(v)/2$ and in strict majority assignment $\tau(v)=\lceil (deg(v)+1)/2 \rceil$. Dynamic monopolies with strict majority thresholds have been studied in \cite{ABW,CL3,C,DR,FKRRS,KSZ,KSZ2}.

\noindent The paper is organized as follows. In Section $2$ we study the spread of influence in weighted undirected and directed networks. Cascade of defaults in banking systems is one main example. Then we introduce dynamic monopolies in weighted graphs and directed graphs and obtain some upper bound for the minimum size of dynamic monopolies. Some bounds are obtained for the strict majority dynamic monopolies. In Section $3$ we obtain an algorithm for the weighted version of ``Target Set Selection" problem with time complexity in terms of the treewidth of graphs. The problem in some special cases has linear time complexity.

\section{Spread of weighted influences}

\noindent In most real world networks the influence is weighted in the sense that unilateral (or bilateral) influence of a vertex to each of its neighbors $w$ depends on the weight of edge between $v$ and $w$. For example in propagation of beliefs or adaptation of new products by ``word of mouth", experts are more influential than ordinary people. Kempe et al.~\cite{KKT} investigate applications of the spread of weighted influence in viral marketing. A well-known network in which influences between nodes are weighed is spread of default or bankruptcy in banking networks. We discuss this network with some details after presenting formal concepts.

\noindent In the above-mentioned real world examples the activation processes depend on the weights or strengths of links (edges). In this paper by $(G, w,\tau)$ we mean any undirected graph without loops or multiple edges together with a weight function $w:E(G)\rightarrow [0, \infty)$ and a threshold assignment $\tau:V(G)\rightarrow [0, \infty)$. An activation process in $(G, w, \tau)$ is defined as follows. Assume that initially some subset $D$ of vertices in $G$ are active. Set $D_0=D$ and denote by $D_i$ the set of active vertices in phase $i$. Then a vertex $u\in V(G)\setminus D_i$ becomes active in phase $i+1$ if and only if the following inequality holds, where $E_i(u)$ consists of all edges say $e$ such that $e=uv$ for some vertex $v\in D_i$. $$\sum_{e\in E_i(u)} w(e) \geq \tau(u).$$

\noindent A dynamic monopoly in $(G, w, \tau)$ is any subset of vertices such that if they get activated at the initial phase then the activation spreads to the whole graph. By $dyn(G, w, \tau)$ we mean the smallest cardinality of dynamic monopolies in the graph. By $(\vec{G}, w, \tau)$ we mean any directed graph $\vec{G}$ in which for any two vertices $u$ and $v$ there exist at most one directed edge from $u$ to $v$. Also $w$ is a weight function on the edges of $\vec{G}$ and $\tau$ a threshold assignment for the vertices of $\vec{G}$. In the activation model for $(\vec{G}, w, \tau)$ assume that each vertex is influenced only by its in-neighbors. Dynamic monopolies and $dyn(\vec{G}, w, \tau)$ are defined similarly for $dyn(\vec{G}, w, \tau)$.

\noindent One of the real world networks realizing the activation model with weighted influence, is the ``spread of default in banking networks". Because this is an important and widely studied topic, we discuss it with some details. According to Cont et al. \cite{CMS}, counterparty relations in financial systems may be represented as a weighted directed graph, defined as a triplet
$I = (V, E, c)$, consisting of a set $V$ of $n$ financial institutions,
a matrix $E$ of bilateral exposures, where $E_{ij}$ represents the exposure of node
$i$ to node $j$ defined as the market value of all liabilities of
institution $j$ to institution $i$ at the date of computation. It is thus the
maximal short term loss of $i$ in case of an immediate default of $j$.
Also $c = (c(i) : i \in V)$, where $c(i)$ is the capital of the institution $i$,
representing its capacity for absorbing losses.
Default occurs when an institution fails to fulfill a legal obligation such as
a scheduled debt payment of interest or principal, or the inability to service
a loan. When a financial institution (say, $c$) defaults, it leads to an immediate
writedown in value of all its liabilities to its creditors. The resulting ``loss cascade" is defined as follows. Consider an initial configuration of capital
reserves $(c(j), j \in V)$. Define the sequence $(c_{k} (j), j \in V)_{k\geq 0}$ as $c_0(j)=c(j)$ and $$c_{k+1}(j)= {\max} \{c_0(j) - \sum_{i: c_k(i)=0} (1-R_i)E_{ji},0\},$$
\noindent where $R_i$ is the recovery rate at the default of institution $i$. Note that $(c_{n-1}(j),j\in V)$ represents the remaining capital once all counterparty losses have been accounted for. The set of insolvent institutions is then given by $D(c,E)= \{j\in V: c_{n-1}(j)=0\}$. We check that a vertex (i.e. an institution) $j$ becomes insolvent during the spread of default if for some $k$, $c_k(j)=0$. According to the model this occurs if $${\max} \{c_0(j) - \sum_{i: c_{k-1}(i)=0} (1-R_i)E_{ji},0\}=0$$
\noindent or equivalently $c_0(j) \leq {\sum}_{i: c_{k-1}(i)=0} (1-R_i)E_{ji}$. In the left side $c_o(j)=c(j)$ is the initial threshold of the vertex $j$ and on the right side we have the total weighted influence which enters to $j$ from the in-neighbors of $j$ which are active in phase $k-1$. Hence the features of the default cascade coincide with our ``threshold based" activation model for weighted graphs. We obtain the following remark.

\begin{remark}
Default contagion in banking systems can be formulated by activation process
in weighted directed graphs representing the network of the system.
\end{remark}

\noindent In the following we prove that the model for weighted graphs is equivalent to an appropriate model in which the underlying graph is not weighted but is multigraph, i.e. a graph with parallel edges. Let ${\mathcal{M}}$ be a multigraph, where between any two vertices $u, v$ there exist $m_{uv}$ parallel edges. In case that there exists no edge between $u$ and $v$ then set $m_{uv}=0$. In multigraphs, spread of influence and dynamic monopolies are defined similarly. In fact, let $D_i$ consist of the active vertices in phase $i$ of the activation process. Then an arbitrary vertex $u\in V({\mathcal{M}})\setminus D_i$ becomes active in phase $i+1$ if and only if ${\sum}_{v\in D_i} m_{uv} \geq \tau(u)$.

\begin{prop}
Corresponding to any $(G, w, \tau)$ there exists a multigraph $\mathcal{M}$ with $V({\mathcal{M}})=V(G)$ and a threshold assignment $\tau'$ such that $D \subseteq V(G)$ is dynamic monopoly in $(G, w, \tau)$ if and only if $D$ is dynamic monopoly in $(\mathcal{M}, \tau')$.\label{weight-multi}
\end{prop}

\noindent\begin{proof}
Without loss of generality we may assume that any weight in $G$ is a non-negative rational number say $p/q$ such that $(p,q)=1$.
Let $\ell$ be the least common divisor of all denominators in $\{w(e):e\in E(G)\}$. Let $e$ be any arbitrary edge of weight $p/q$ between say $u$ and $v$ in $G$, where $(p,q)=1$. Replace $e$ by $\ell \times (p/q)$ parallel edges between $u$ and $v$. For any vertex $u$ in $G$, replace its threshold $\tau(u)$ by the new threshold $\ell \tau(u)$. Denote the resulting multigraph and the new threshold assignment by $\mathcal{M}$ and $\tau'$, respectively. Assume that $v$ is an arbitrary vertex in $G$ and $e_1, \ldots, e_k$ are some edges incident to $v$ such that $w(e_1)+ \cdots + w(e_k) \geq \tau(v)$. Then ${\sum}_i \ell w(e_i) \geq \ell \tau(v)$. Let $e_i=vu_i$. Note that $\ell w(e_i)$ is the multiplicity of the edge $vu_i$ in the multigraph ${\mathcal{M}}$. It follows that ${\sum}_i m_{vu_i} \geq \tau'(v)$. Reverse of the above inequalities also hold. This argument asserts that if beginning from any initially active subset $D$ of vertices in $(G, w, \tau)$, a vertex $v$ becomes active in a phase say $j$ in $(G, w, \tau)$ then $v$ becomes active at phase $j$ in $(\mathcal{M}, \tau')$ provided that $D\subseteq \mathcal{M}$ is activated at phase 0 in $\mathcal{M}$. The converse also holds. It implies that $D$ is dynamic monopoly in $(G, w, \tau)$ if and only if $D$ is dynamic monopoly in $(\mathcal{M}, \tau')$.
\end{proof}

\noindent For $(\vec{G}, w, \tau)$ we have the following analogous result with a proof similar to that of Proposition \ref{weight-multi}.

\begin{prop}
Corresponding to any $(\vec{G}, w, \tau)$ there exists a directed multigraph $\vec{\mathcal{M}}$ with $V({\vec{\mathcal{M}}})=V(\vec{G})$ and a threshold assignment $\tau'$ such that $D \subseteq V(\vec{G})$ is dynamic monopoly in $(\vec{G}, w, \tau)$ if and only if $D$ is dynamic monopoly in $(\vec{\mathcal{M}}, \tau')$.\label{weight-directed-multi}
\end{prop}

\noindent In any simple graph $G$ or multigraph ${\mathcal{M}}$, by a vertex cover we mean any subset $S$ of vertices in $G$ (resp. ${\mathcal{M}}$) such that any edge of $G$ (resp. ${\mathcal{M}}$) has at least one endpoint in $S$. Following \cite{W}, we denote by $\beta(G)$ the minimum cardinality of any vertex cover in $G$. Similarly, denote by $\beta({\mathcal{M}})$ the minimum cardinality of any vertex cover in ${\mathcal{M}}$.

\begin{prop}
For any $(G, w, \tau)$, $dyn(G, w, \tau)\leq \beta(G)$.
\end{prop}

\noindent\begin{proof}
Let $\mathcal{M}$ and $\tau'$ be the multigraph and threshold assignment corresponding to $(G, w, \tau)$ as indicated by Proposition \ref{weight-multi}. It is enough to prove that $(\mathcal{M}, \tau')$ has a dynamic monopoly with no more than $\beta(G)$ vertices. Note that $V(G)=V(\mathcal{M})$. Let $S$ be a minimum vertex cover in $\mathcal{M}$. We show that $S$ is a $\tau'$-dynamic monopoly in $\mathcal{M}$. Let $u\in V(\mathcal{M})\setminus S$. Since $S$ is vertex cover then any neighbor of $u$ belongs to $S$. Hence at least $\tau'(u)$ neighbors of $u$ are in $S$. It follows that $S$ is a dynamic monopoly. This completes the proof.
\end{proof}

\noindent It was proved in \cite{ABW} that any graph $(G, \tau)$ admits a $\tau$-dynamic monopoly with at most ${\sum}_{v\in G} \tau(v)/(1+deg(v))$ vertices. We use the proof idea of this bound and generalize it for multigraphs. Then in the light of Proposition \ref{weight-multi} we obtain the following bound for weighted graphs.
Note that the degree of each vertex $u$ in a multigraph ${\mathcal{M}}$ is defined as $d(u)={\sum}_{v\in {\mathcal{M}}} m_{uv}$. In the following theorem by $d(v)$ we mean the degree of a vertex $v$ in a weighted graph $G$. Also $N(v)$ stands for the neighborhood set of $v$.

\begin{prop}
Let $(G, w, \tau)$ be given. Then there exists a dynamic monopoly $D$ in $(G, w, \tau)$ such that $$|D|\leq \sum_{v\in G}~~~ \sum_{S\subseteq N(v): \sum_{u\in S} m_{uv} < \ell \tau(v)} \frac{|S|!(d(v)-|S|)!}{(d(v)+1)!}.$$\label{upper-bound}
\end{prop}

\noindent\begin{proof}
Let $\ell$ be the least common divisor of all denominators in $\{w(e):e\in E(G)\}$. Let ${\mathcal{M}}$ with $V(G)=V({\mathcal{M}})$ be the multigraph as constructed in proof of Proposition \ref{weight-multi}. Set $\tau'(v)=\ell \tau(v)$ for each vertex $v$ in ${\mathcal{M}}$. Let $v_1, \ldots, v_n$ be a random list of vertices in ${\mathcal{M}}$. It is easily observed that for any such list the set $D=\{v_i: deg_{\{v_1, \ldots, v_{i-1}\}}(v_i)<\tau'(v_i)\}$ is a $\tau'$-dynamic monopoly in ${\mathcal{M}}$. We claim that $$Pr(deg_{\{v_1, \ldots, v_{i-1}\}}(v_i)< \tau'(v_i)) = \sum_{S\subseteq N(v_i): {\sum}_{u\in S} m_{uv_i} < \ell \tau(v_i)} \frac{|S|!(d(v_i)-|S|)!}{(d(v_i)+1)!}.$$
\noindent For this purpose, first note that the vertices which are not neighbor of $v_i$ are irrelevant in $Pr(deg_{\{v_1, \ldots, v_{i-1}\}}(v_i)< \tau'(v_i))$. In fact, we should consider $(d(v_i)+1)!$ random lists of vertices in $N(v_i)\cup \{v_i\}$ and determine the probability that at most $\tau'(v_i)$ edges exists between $v_i$ and those vertices which appear before $v_i$ in the list. By the ordinary counting method we obtain that the probability that there exists exactly $j$ edges between $v_i$ and the vertices in the list which are before $v_i$ equals $$\sum_{S\subseteq N(v_i): \sum_{u\in S} m_{uv_i} =j} \frac{|S|!(d(v_i)-|S|)!}{(d(v_i)+1)!}.$$
\noindent Hence the claim is proved. It follows that
$$\Bbb{E}(|D|)\leq \sum_{v\in {\mathcal{M}}}~~ \sum_{S\subseteq N(v): {\sum}_{u\in S} m_{uv} < \ell \tau(v)} \frac{|S|!(d(v)-|S|)!}{(d(v)+1)!}.$$
\noindent Hence, there exists a $\tau'$-dynamic monopoly in ${\mathcal{M}}$ with at most the previously mentioned number of vertices. The proof is completed by applying Proposition \ref{weight-multi} for $G$ and ${\mathcal{M}}$.
\end{proof}

\noindent It's clear that if in the bound of Proposition \ref{upper-bound}, $N(v)$ is replaced by the set of in-neighbors of $v$, then the same bound holds for directed graphs. In the following we obtain some bounds for the strict majority dynamic monopolies in multigraphs/directed multigraphs and weighted graphs or directed graphs. In a multigraph ${\mathcal{M}}$ the strict majority threshold is naturally defined as $\tau(u)=\lceil (deg(u)+1)/2 \rceil$ for any vertex $u$, where $deg(u)= {{\sum}}_{v\in {\mathcal{M}}} m_{uv}$. When we go to a weighted graph $(G,w)$, since the edges have not a same weight then the strict majority threshold cannot be defined as $\lceil (deg_G(u)+1)/2 \rceil$. How is the strict majority threshold defined for weighted graphs $(G, w)$? To obtain the answer we should find a common module for the weight of edges in $(G,w)$. Since each edge of weight $p/q$ is replaced by $\ell \times (p/q)$ parallel edges with unit weight in ${\mathcal{M}}$ then $1/\ell$ is a common module for all weighted edges in $(G,w)$. Hence the strict majority threshold in $(G,w)$ is defined as $\tau(v)= \lfloor ({\sum}_{e: v\in e} w(e)) /2 + (1/\ell) \rfloor$. This topic is easily generalized for weighted directed graphs. We obtain the following remark.

\begin{remark}
Let $(G,w)$ (resp. $(\vec{G},w)$) be any weighted graph (resp. directed graph) and ${\mathcal{M}}$ (resp. $\vec{{\mathcal{M}}}$) be its corresponding multigraph (resp. directed multigraph). Then any strict majority dynamic monopoly for ${\mathcal{M}}$ (resp. $\vec{{\mathcal{M}}}$) is a strict majority dynamic monopoly for $(G,w)$ (resp. $(\vec{G},w)$) and vice versa.\label{rem1}
\end{remark}

\noindent Upper bounds for the strict majority dynamic monopolies of simple graphs were obtained by Khoshkhah et al. in \cite{KSZ}. In Theorem \ref{bound-major-multi} we prove that any multigraph ${\mathcal{M}}$ on $n$ vertices admits a strict majority dynamic monopoly with no more than $\lceil n/2 \rceil$ vertices. By generalizing a function introduced in \cite{KSZ}, we first define the following function for multigraphs. Let ${\mathcal{M}}$ be a multigraph on $n$ vertices. By an ordering $\sigma$ on the vertex set of ${\mathcal{M}}$, we mean any bijective function $\sigma:V({\mathcal{M}})\rightarrow \{1, 2, \ldots, n\}$. Let $\sigma$ be an ordering on the vertex set of ${\mathcal{M}}$. The function $f_{\sigma}:V({\mathcal{M}})\rightarrow \Bbb{Z}$ is defined as follows for any vertex $v$ of ${\mathcal{M}}$:
$$f_{\sigma}(v)= \sum_{u: \sigma (u)> \sigma (v)} m_{uv} ~-~ \sum_{u: \sigma (u) < \sigma (v)} m_{uv}.$$

\begin{thm}
Let ${\mathcal{M}}$ be a multigraph on $n$ vertices. Then there exists a strict majority dynamic monopoly $D$ for ${\mathcal{M}}$ with at most $\lceil n/2 \rceil$ vertices. Moreover, there exists an ${\mathcal{O}}(n^2)$ algorithm which outputs such a set $D$.\label{bound-major-multi}
\end{thm}

\noindent \begin{proof}
Let $\sigma$ be an arbitrary ordering on the vertices of ${\mathcal{M}}$.
Define $D_1=\{v: f_{\sigma}(v)\geq 0\}$. We prove that $D_1$ is a strict majority dynamic monopoly. In fact the vertices with negative $f$ become active in turn according to their order in $\sigma$. For this purpose let $w$ be the first vertex with $f_{\sigma}(w)<0$. Hence each vertex $v$ with $\sigma(v)<\sigma(w)$ satisfies $f_{\sigma}(v)\geq 0$ and then $v\in D_1$. Since $f_{\sigma}(w)<0$ then
$$\sum_{u: \sigma (u)> \sigma (w)} m_{uw} \leq (\sum_{u: \sigma (u) < \sigma (w)} m_{uv})-1.$$
\noindent We have $deg(w)={\sum}_{u: \sigma (u)> \sigma (w)} m_{uw} +  {\sum}_{u: \sigma (u) < \sigma (w)} m_{uw}$. It follows that $$\frac{deg(w)+1}{2} \leq \sum_{u: \sigma (u) < \sigma (w)} m_{uw} \leq deg_{D_1}(w).$$ It implies that $w$ becomes active in phase $1$. Let $w'$ be the first vertex after $w$ with $f_{\sigma}(w')<0$. A similar argument shows that $w'$ has at least $(deg(w')+1)/2$ neighbors in $D_1\cup \{w\}$. Continuing this technique eventually all vertices in ${\mathcal{M}}$ becomes active.

\noindent Define $D_2 = \{v: f_{\sigma}(v)\leq 0\}$. A similar argument shows that $D_2$ is a dynamic monopoly. In fact the vertices with positive $f$ become active in turn according to reverse of their order in $\sigma$. Now either $|D_1| \leq \lceil n/2 \rceil$ or $|D_2| \leq \lceil n/2 \rceil$. This proves the bound of the theorem. Note that to specify $D_1$ and $D_2$, ${\mathcal{O}}(n^2)$ sums and comparisons is enough. This completes the proof.
\end{proof}

\noindent The following results are immediate from Remark \ref{rem1}. The required dynamic monopolies are obtained in polynomial time.

\begin{cor}
For any $(G, w, \tau)$, where $\tau$ is the strict majority threshold $$dyn(G, w, \tau) \leq \lceil |G|/2 \rceil.$$
\end{cor}

\begin{remark}
For any $(\vec{G}, w, \tau)$, where $\tau$ is the strict majority threshold $$dyn(\vec{G}, w, \tau) \leq \lceil |\vec{G}|/2 \rceil.$$
\end{remark}

\section{Relationships with treewidth}

\noindent Treewidth of graphs is a very useful concept in algorithmic study of graph theoretical problems. Treewidth and tree decomposition of graphs have a few equivalent definitions. We use the one introduced in the textbook \cite{D}. A tree decomposition of a graph $G$ is a tree $T$, with nodes $X_1, \ldots, X_n$, where each $X_i$ is a subset of $V(G)$, satisfying the following properties:

\noindent (1) ${\bigcup}_i X_i=V(G)$.

\noindent (2) If $X_i$ and $X_j$ both contain a vertex $v$, then all nodes $X_k$ of $T$ in the path between $X_i$ and $X_j$ contain $v$ as well.

\noindent (3) For every edge $uv$ in the graph, there is a subset $X_i$ that contains both $u$ and $v$.

\noindent The width of a tree decomposition is the size of its largest set $X_i$ minus one. The treewidth $tw(G)$ of $G$ is the minimum width among all possible tree decompositions of $G$.

\noindent Let $(G,w)$ be any simple weighted graph, where all weights are rational numbers of the form say $p/q$ with $(p,q)=1$. Let $\ell$ be the least common divisor of all denominators in $w$. Obtain a simple (non-weighted) graph $H=H(G,w)$ from $(G,w)$ as follows. First, in the graph $G$ replace each edge $e=uv$ with weight $p/q$ by $\ell p/q$ parallel edges between $u$ and $v$. Then in case that $\ell p/q \geq 2$ replace $\ell p/q -1$ many of parallel edges between $u$ and $v$ by a path of length two between $u$ and $v$. Denote the resulting graph by $H$.

\begin{lemma}
Let $(G,w)$ be any simple weighted graph, where all weights are rational number of the form say $p/q$ with $(p,q)=1$. Let $\ell$ be the least common divisor of all denominators in $w$. Then for $H=H(G,w)$ we have $$tw(H)\leq {\max} \{tw(G), {\max} \{\ell w(e): e\in E(G)\}\}.$$\label{width-lem}
\vspace*{-0.5cm}
\end{lemma}

\noindent \begin{proof}
Set $tw(G)=t$ and let $T$ be a tree decomposition of $G$ with node sets $X_1, \ldots, X_n$ such that ${\max}_{i=1}^n |X_i| -1=t$. We obtain a tree decomposition for $H$ from $T$. Let $u$ and $v$ be any two adjacent vertices in $G$. Let the edge $uv$ has weight $w(uv)$. If $\ell w(uv)=1$ then there is no vertex of $H$ between $u$ and $v$. But when $\ell w(uv)\geq 2$ then there are $\ell w(uv)-1$ vertices of $H$ say $w_1, \ldots, w_k$ between $u$ and $v$. We construct a new node set $X_{uv}$ consisting of $u$ and $v$ and also $w_1, \ldots, w_k$. Since $u$ and $v$ are adjacent there is a node set say $X_j$ in $T$ which contains both $u$ and $v$. Put an edge between $X_j$ and $X_{uv}$. Note that $|X_{uv}|=\ell w(uv)+1$. Do this operation for each edge $uv$ of $G$ satisfying $\ell w(uv)\geq 2$. We obtain a tree decomposition $T'$ for $H$. It is easily observed that each node set in $T'$ has either at most $t+1$ vertices or at most ${\max}_{e\in E(G)} \ell w(e)+1$ vertices. It follows that $tw(H)\leq {\max} \{tw(G), {\max}_{e\in E(G)} \ell w(e) \}$.
\end{proof}

\noindent Let $(G,w,\tau)$ be any weighted graph
and set $H=H(G,w)$. Define a threshold assignment $\tau'$ for the vertices of $H$ as follows. Recall that for each edge $uv$ in $G$ such that $\ell w(uv)\geq 2$ we have $\ell w(uv) - 1$ middle vertices of degree two between $u$ and $v$ in $H$. For all these middle vertices $w$ define $\tau'(w)=1$. For each vertex $u\in V(G)$ set $\tau'(u)=\ell \tau(u)$. We have the following proposition.

\begin{prop}
$$dyn(G, w, \tau)=dyn_{\tau'}(H).$$\label{prop-width}\vspace{-0.3cm}
\end{prop}
\noindent \begin{proof}
Let $D$ be any dynamic monopoly for
$(G, w, \tau)$. Note that each middle vertex in $H$ is adjacent to some vertices in $G$ and has threshold $1$. Hence after activation of all vertices of $G$ then the middle vertices get activated too. It implies that $D$ is a $\tau'$-dynamic monopoly for $H$ and $dyn_{\tau'}(H)\leq dyn(G, w, \tau)$. To prove the converse of inequality assume that $M$ is any $\tau'$-dynamic monopoly for $H$ with minimum cardinality. There are two possibilities for the vertices of $M$.

\noindent {\bf Case 1:} $M$ contains no middle vertices of $H$. In this case $M\subseteq V(G)$. Activation of any vertex $u$ of $M$ instantly activates all possible middle vertices which are neighbors of $u$. It implies that $M$ is a dynamic monopoly for $(G, w, \tau)$.

\noindent {\bf Case 2:} In this case assume that $M$ contains a middle vertex say $w$ which is between $u$ and $v$ from $G$. Since the threshold of $w$ is one then neither $u$ nor $v$ belongs to $M$. Otherwise $M\setminus \{w\}$ is $\tau'$-dynamic monopoly which contradicts the minimality of $M$. Now, define a new set $M'=M\setminus \{w\} \cup \{u\}$. When the vertices of $M$ are initially activated, $w$ becomes active at the next round of the activation because the threshold of $w$ is one. It follows that $M'$ is a minimum dynamic monopoly for $(H, \tau')$ such that the number of middle vertices in $M'$ is strictly less than that of $M$. By continuing this method we eventually obtain a minimum dynamic monopoly $M''$ of $(H,\tau')$ which satisfies the case $1$ above. This completes the proof.
\end{proof}

\noindent Let $T$ be any tree graph and $\tau$ a
threshold assignment for the vertices of $T$.
Let $u$ be any vertex of degree one in $T$ and $w$ its unique neighbor in $T$. It is easily seen that there exists a minimum dynamic monopoly for $T$ which does not contain $u$. Since let $D'$ be a minimum dynamic monopoly containing the vertex $u$. Then $D=D'\setminus \{u\} \cup \{w\}$ is a minimum dynamic monopoly for $T$. This fact is the base of a polynomial time algorithm to determine $dyn_{\tau}(T)$. In fact we remove $u$ from $T$ and keep the threshold of $w$ unchanged. Then we repeat this technique for the rest of graph. Note that trees are graphs with treewidth one. This technique can be generalized in sophisticated form for graphs with bounded treewidth. In fact the authors of \cite{BHLN} prove that the Target Set Selection problem for input graphs $G$ on $n$ vertices and treewidth $tw(G)$ can be solved by an algorithm with time complexity ${\mathcal{O}(n^{tw(G)})}$. For weighted graphs $(G,w)$ we introduce the following quantity as treewidth of $G$. For any $(G, w, \tau)$ define $tw(G, w)={\max}\{tw(G), {\max}\{\ell w(e):e\in E(G)\}\}.$

\begin{thm}
For any input $(G, w, \tau)$ with the corresponding value $\ell$ denote ${\max}_{e\in E(G)} \ell w(e)$ by $\mu=\mu(G)$. There exists an algorithm with time complexity ${\mathcal{O}((n+m\mu)^{tw(G,w)})}$ which given as input $(G, w, \tau)$ on $n$ vertices and $m$ edges, returns a minimum dynamic monopoly for $(G, w, \tau)$.\label{width-weight}
\end{thm}

\noindent \begin{proof}
By Proposition \ref{prop-width}, $dyn(G, w, \tau)=dyn_{\tau'}(H)$, where $H=H(G,w)$. 	
By Lemma~\ref{width-lem}, $tw(H)\leq {\max} \{tw(G), {\max} \{\ell w(e): e\in E(G)\}\}$.
Note that $$|V(H)|=|V(G)|+\ell \sum_{e\in E(G)}w(e)\leq |V(G)|+|E(G)|\mu(G).$$
\noindent Now, by applying the algorithm of \cite{BHLN} for $H$ we obtain $dyn_{\tau'}(H)$ with a time complexity mentioned in the theorem.
\end{proof}


\noindent Let ${\mathcal{F}}$ be the family consisting of directed graphs $\vec{G}$ whose vertices can be ordered as $v_1, \ldots, v_n$ such that the in-degree of $v_i$ in $\vec{G}[v_1, \ldots, v_i]$ (i.e. $d^{in}_{\vec{G}[v_1, \ldots, v_i]}(v_i)$) is at most one.

\begin{thm}
There exists a linear time algorithm which given any input $(\vec{G},w,\tau)$ with $\vec{G}\in {\mathcal{F}}$ outputs a dynamic monopoly with minimum size for $(\vec{G}, w, \tau)$.\label{1-dir-degen}
\end{thm}

\noindent \begin{proof}
Let $v_1$ be a vertex of in-degree one in $\vec{G}$ and $u$ be its in-neighbor. We claim that there exists a minimum dynamic monopoly $D$ for $(\vec{G}, w, \tau)$ such that $v_1\not\in D$. Let $D'$ be any minimum dynamic monopoly such that $v_1\in D_1$. Then $u\not\in D'$ since $D'$ is minimal. Hence $D=D'\setminus \{v_1\} \cup \{u\}$ is a minimum dynamic monopoly for $(\vec{G}, w, \tau)$ with $v_1\not\in D$. For $\vec{G}\setminus v_1$, define a threshold assignment $\tau_0$ obtained by restricting $\tau$ to the vertices of $\vec{G}\setminus v_1$. Let also $w_0=w {\mid}_{E(\vec{G}\setminus v_1)}$. We have now $dyn(\vec{G}, w, \tau)=dyn(\vec{G}\setminus \{v_1\}, w_0, \tau_0)$. This equality is the base of our recursive algorithm to gradually obtain a set $D$ which will be a dynamic monopoly for $(\vec{G}, w, \tau)$. In $\vec{G}\setminus v_1$ there are two possibilities for $\tau(u)=\tau_0(u)$.

\noindent {\bf Case 1.} If $\tau(u)> d^{in}_{\vec{G}\setminus v_1}(u)$ then we put $u$ in a set $D$. Then remove $u$ from $\vec{G}\setminus v_1$ and set $\vec{G'}=\vec{G}\setminus \{v_1,u\}$ and also decrease the threshold of each vertex in $\vec{G'}$ by one. Let $\tau'$ be the resulting threshold assignment. Denote by $w'$ the restriction of $w$ to $E(\vec{G'})$. Note that in this case any dynamic monopoly for $(\vec{G}\setminus \{v_1\}, w_0, \tau_0)$ necessarily contains $u$. It follows that $dyn(\vec{G}, w, \tau)=dyn(\vec{G'}, w', \tau')+1$. Since $\vec{G'}\in {\mathcal{F}}$ we repeat the same technique for $(\vec{G'}, w', \tau')$.

\noindent {\bf Case 2.} If $\tau(u)\leq d^{in}_{\vec{G}\setminus v_1}(u)$ then $\vec{G}\setminus v_1$ contains a vertex say $v_2$ of in-degree one. We keep the set $D$ unchanged and then we repeat the same technique for $\vec{G}\setminus v_1$ and $v_2$.

\noindent The above-mentioned technique outputs recursively a set $D$ which is a dynamic monopoly for $\vec{G}$. Note that the total number of steps is $\mathcal{O}(|V(\vec{G})|)$.
\end{proof}

\noindent The undirected version of Theorem \ref{1-dir-degen} also holds with a similar proof. First time Chen~\cite{C} obtained a linear time algorithm for determining the smallest target set in trees. The following remark extends Chen's result for trees with weighted edges.

\begin{remark}
Target Set Selection can be solved in linear time for edge-weighted trees.
\end{remark}

\section{Concluding remarks}

\noindent In Theorem \ref{width-weight} we obtained an ${\mathcal{O}((n+m\mu)^{tw(G,w)})}$ algorithm for the target set selection problem for input graphs $G$ with $n$ vertices and $m$ edges, where $\mu= \ell {{{\max}}_{e\in E(G)}} w(e)$. It is an interesting challenge to obtain an algorithm which accomplishes the same job but with a time complexity ${\mathcal{O}(n^{f(tw(G,w))})}$, where $f(tw(G,w))$ is a function only in terms of $tw(G,w)$.

\section{Acknowledgment}

\noindent The author thanks the anonymous referee for kindly refereeing the paper.

\end{document}